\documentclass[prb,aps,twocolumn,superscriptaddress,showpacs]{revtex4}
\usepackage{graphicx}
\usepackage{psfrag}
\usepackage{color}
\usepackage{subfigure}
\usepackage{amsmath}
\definecolor{dred}{rgb}{0.7,0.0,0.0}
\allowdisplaybreaks 
 
\begin{document}

%
%

\title{Pairing Symmetries of a Hole-Doped Extended Two-Orbital Model for 
the Pnictides}

\author{Andrew Nicholson}
\author{Weihao Ge}
 
\affiliation{Department of Physics and Astronomy, The University of
  Tennessee, Knoxville, TN 37996} 
\affiliation{Materials Science and Technology Division, Oak Ridge
  National Laboratory, Oak Ridge, TN 37831} 

\author{Jos\'e Riera}

\affiliation{Instituto de F\'isica Rosario, Universidad Nacional de Rosario, 2000-Rosario, Argentina}


\author{Maria Daghofer}

\affiliation{IFW Dresden, P.O. Box 27 01 16, D-01171 Dresden, Germany}





\author{Adriana Moreo}
\author{Elbio Dagotto}

\affiliation{Department of Physics and Astronomy, The University of
  Tennessee, Knoxville, TN 37996} 
\affiliation{Materials Science and Technology Division, Oak Ridge
  National Laboratory, Oak Ridge, TN 37831}

\date{\today}

\begin{abstract}
The hole-doped ground state of a recently introduced extended ``$t$-$U$-$J$'' 
two-orbital Hubbard model for the Fe-based superconductors is studied via
exact diagonalization methods on small clusters. 
Similarly as in the previously studied case of electron doping,
A. Nicholson {\it et al.}, Phys. Rev. Lett. {\bf 106} 21702 (2011),
upon hole doping it is observed that 
there are several competing pairing symmetries including $A_{1g}$,
$B_{1g}$, and $B_{2g}$. However, contrary to the electron-doped case, 
the ground state of the hole-doped state 
has pseudocrystal momentum ${\bf k}=(\pi,\pi)$ in the unfolded Brillouin zone. 
In the two Fe-atom per unit cell
representation, this indicates that the ground state involves anti-bonding, 
rather than bonding, combinations of the orbitals of the two Fe atoms in the
unit-cell. The lowest state with ${\bf k}=(0,0)$ has only a slightly higher energy. 
These results indicate that this simple two-orbital model may be useful to 
capture some subtle aspects of the hole-doped pnictides 
since calculations for the five-orbital model have unveiled a hole pocket
centered at $M$ (${\bf k}=(\pi,\pi)$) in the unfolded Brillouin zone.


\end{abstract}

\maketitle

\section {Introduction} 

The detailed study of the 
recently discovered high critical temperature superconductivity (HTCS) in 
the iron-based pnictides and chalcogenides\cite{johnston} 
continues providing important information to understand 
the still puzzling mechanism that drives this remarkable phenomenon. Experiments indicate
that these Fe-based materials share many properties with the high $T_c$ 
cuprates,\cite{dagottoRMP} such as magnetically ordered parent 
compounds\cite{dai} and superconducting states stabilized by 
either electron or hole doping.\cite{johnston} However, there are also 
remarkable differences, such as the fact that the parent compounds, 
at least for the pnictides, are 
(bad) metals rather than Mott insulators. Moreover, several of the iron
$d$-orbitals are active at the Fermi surface (FS) as opposed to the case of
the cuprates where just the copper $d_{x^2-y^2}$ orbital 
plays the major role. While it
has been clearly established that the superconducting pairing operator has
$d$-wave symmetry in the case of the hole-doped cuprates,\cite{ibm} the symmetry
of the pairing operator in the pnictides is still controversial: surface-sensitive 
angle-resolved photoemission (ARPES) studies~\cite{arpes}
indicate that full nearly-momentum-independent gaps open on all FS pockets, 
compatible with the $s_{\pm}$ state.~\cite{teo} However, several other 
experiments testing bulk properties provide 
results compatible with nodal superconductivity.~\cite{nodal}

Reliable theoretical studies are difficult to implement for the complex 
multi-orbital models needed for these materials
without making explicit assumptions about the ground state properties 
or about the mechanism and strength of the  pairing 
interactions. Under the assumption of a magnetically driven 
superconducting instability, the
Random Phase Approximation (RPA) is often applied in this context 
providing indications that
several pairing channels, mostly $A_{1g}$ and $B_{1g}$, are in 
competition in this type of compounds.\cite{graser} 
However, RPA relies on a particular subset
of Feynman diagrams and it is a weak-coupling approach. On the opposite extreme,
strong coupling studies have
also been performed. Depending on the particular model used, some authors have found 
evidence of pairing with $B_{2g}$ symmetry \cite{si} while others have found 
a variety of competing states.\cite{berne} A complementary approach to these
previous investigations is to perform an exact diagonalization (ED) of the
model Hamiltonians, allowing
to solve the problem exactly for any value of the interaction. However, since the
Hilbert space grows exponentially with the system size, this method can be implemented only in very small
clusters and with a reduced number of active Fe orbitals. This ED approach has been
recently applied by the authors and collaborators 
to the study of an electron-doped two-orbital Hubbard 
model and in that effort the presence of competing pairing states was observed 
as the strength of the coupling parameters was varied.\cite{andrew,Daghofer:2008,moreo}

An important characteristic of the widely studied Hubbard 
models for the pnictides/chalcogenides is that they are
not particle-hole symmetric. On the experimental side, superconductivity has
been found both upon electron and hole doping, but it seems that hole-doped
materials belonging to the 122 family are more suitable for the use of 
surface-sensitive techniques,
while electron-doped materials belonging 
to the 1111 family are more easily studied with 
bulk techniques.\cite{arpes,nodal} 
Then, it is natural to wonder whether a potential source of the 
differences in the experimental results regarding the pairing symmetries may 
arise from the nature of the dopants. For this reason it is important to study 
theoretically the properties of multi-orbital Hubbard models both under 
electron and hole doping. Previously, such a comparative analysis has been performed employing the RPA method applied to a five-orbital Hubbard 
model.\cite{kemper} For the parameter range 
studied in that case (weak coupling), a pairing state with $A_{1g}$ symmetry
was observed in both cases. The state found has nodes on the electron pockets in
the electron-doped case, but upon hole-doping an extra hole pocket at the 
$M=(\pi,\pi)$ point in the Brillouin zone 
leads to the removal of the nodes and the development 
of a nodeless $s_{\pm}$ state.\cite{kemper} 

The goal of the present publication 
is to study the most favorable pairing channels of a 
two-orbital Hubbard model using small-cluster 
 exact diagonalization techniques (namely,
the Lanczos algorithm) for the
case of hole doping, 
and to contrast the results against those found for the case of
electron-doping of the same model that have been recently 
reported.\cite{andrew,Daghofer:2008,moreo} To reduce the severe constraints
imposed by the small size of the clusters that can be diagonalized in present
day computers, a simple generalization of the Hubbard 
model for the pnictides, 
previously introduced for the study of the electron doped
case,\cite{andrew} will be here applied. For this purpose, 
Heisenberg ``$J$'' terms will be added
to the original Hubbard model to enhance spin order and
pairing tendencies, but without projecting
out doubly occupied sites and charge fluctuations.
These terms help to establish tightly bound-states upon doping
that can be studied with Lanczos methods on the small clusters currently 
accessible with state-of-the-art computers. Comparisons with RPA results 
for the five-orbital model and with experimental data will be performed.

The organization of the paper is as follows: the model and the method are
reviewed in Section~\ref{sec:model}, the main results for hole doping 
are presented in 
section~\ref{results}, while section~\ref{conclusions} is devoted to the 
conclusions.

\section {Model and Method}\label{sec:model} 

The model studied here is based on the well-known and widely used two-orbital Hubbard
model~\cite{Daghofer:2008,moreo,raghu} that employs the $d_{xz}$ ($x$) and
$d_{yz}$ ($y$) Fe orbitals. These orbitals
provide the largest contribution to the pnictides' band 
structure at the FS.~\cite{phonon} The reduction in the actual number
of active orbitals in the pnictides is necessary in order to perform the present
Lanczos studies. Calculations with more orbitals for the same
cluster studied here are simply
not possible at present.

The parameters of the electronic hopping terms of the model  were 
previously chosen to provide a close agreement with 
the band structure calculations obtained with density-functional
theory.~\cite{raghu} In addition to the hopping terms, the model also includes
the on-site Coulomb interaction consisting of intra- and
inter-orbital Coulomb repulsions with couplings $U$ and $U'$, the Hund's rule coupling
$J_{\rm H}$, and the pair-hopping term with strength $J'$. While $U$ can in principle
depend on the particular orbital due to different screening effects for each
orbital, this is not the case
for the $d_{xz}$ and $d_{yz}$ orbitals that form a degenerate $e_g$ doublet, implying
that the relations $U'$=$U-2J_{\rm H}$, and $J'$=$J_{\rm H}$ are fulfilled
for symmetry reasons.~\cite{oles83} This Hamiltonian, with the
kinetic-energy hopping and onsite-interaction terms above mentioned, has been studied in detail
previously.~\cite{Daghofer:2008,moreo,raghu,rong} Moreover, in a recent investigation
of the electron-doped case,~\cite{andrew} the model was supplemented
by a Heisenberg interactions to amplify the strength of the magnetic state
and, consequently, the pairing strength as well.


Naively, it may seem that selecting a stronger on-site Hubbard
interaction would stabilize a stronger antiferromagnetic state. However, this
procedure also induces an
insulator, and actually the strength of the effective coupling between the
Fe-spins decreases as $1/U$ with increasing $U$. To avoid this problem,
in early studies of the one-band $t$-$U$-$J$ model~\cite{Daul00} and in a recent study of
the electron-doped two-orbital model,~\cite{andrew} Heisenberg terms have
been added and shown to enhance pairing tendencies. 
Since our aim is to investigate the
symmetries of the Cooper pairs, the  additional magnetic
interactions must have the same symmetries as the original
Hamiltonian. To make sure that the symmetries are properly handled,
the additional Heisenberg interaction is given by the
operatorial  form that corresponds to the
superexchange terms derived from the strong-coupling (large-$U$) limit. In
the case of the one-band Hubbard model, this is a Heisenberg term with
spin $S=1/2$. In the case of a multi-orbital model away from
half-filling, the corresponding superexchange contains an orbital degree of freedom
in addition to the spin and it is of a Kugel-Khomskii
type.~\cite{KK82,Kru09} 

In the present case of a half-filled two-orbital model, the low-energy
Hilbert space for the strong-coupling limit, with both $U$ and $J_{\rm H}$ large,
is given by doubly occupied sites with singly occupied orbitals. 
Due to the Hund's coupling, the
two electrons per site form a triplet state, with an energy 
$E_0=U'-J_{\rm H}=U-3J_{\rm H}$, compared to $E_1=U'+J_{\rm
  H}=U-J'=E_0+2J_{\rm H}$ and $E_2=U+J_{\rm H}=E_0+4J_{\rm H}$ for
inter- and intra-orbital singlet states. The low-energy Hilbert space
is, thus, given by a spin $S=1$ at each site. The interaction between
these spins can be obtained by second-order perturbation theory in an analogous
manner as the well-known derivation of the Heisenberg model from the
one-orbital Hubbard model. The calculation for the two orbitals is the most
easily carried out when the hopping term preserves orbital
flavor, because the first hopping process, which creates a virtual
excitation with energy $U+J_{\rm H}$, then has to involve the same
orbital as the second, which goes back to the low-energy Hilbert
subspace. By this procedure it can be shown 
that the result is the isotropic Heisenberg
interaction for $S=1$ with a coupling 
\begin{equation}\label{eq:Jeff}
J_{\rm eff} = \frac{2}{3}\frac{t_a^2+t_b^2}{U+J_{\rm H}}\;,
\end{equation}
where $t_a$ and $t_b$ are the hopping parameters corresponding to the two
orbitals. With the notation $t_{a/b}=t_{1/2}$,~\cite{raghu} the
nearest-neighbor (NN) coupling $J_{\rm NN}$ can be derived. 
For a next-nearest--neighbor (NNN)
coupling, which is natural since in the original Hubbard model 
the hoppings involve both NN and NNN Fe atoms,
it is convenient to transform to a rotated orbital basis
$(|xz\rangle \pm |yz\rangle)/\sqrt{2}$, where the hoppings are diagonal
in orbital space and given by $t_3\pm t_4$, leading to 
\begin{equation}\label{eq:JNNN}
J_{\rm NNN} =
\frac{4}{3}\frac{t_3^2+t_4^2}{U+J_{\rm H}}=2\frac{t_3^2+t_4^2}{t_1^2+t_2^2}J_{\rm NN}\;.
\end{equation}
Similarly as in the previous investigation of Ref.~\onlinecite{andrew} 
and to avoid the proliferation of parameters, 
the ratio $J_{\rm NNN}/J_{\rm NN}$ is kept fixed to 0.93, 
which is the value that results
from $t_1=-1$, $t_2=1.3$, $t_3=t_4=-0.85$.\cite{raghu} For the
electron-doped case, the results were found to be  not sensitive to changes in this
ratio as long as the model remains in the regime with
$(\pi,0)$/$(0,\pi)$ magnetic order.\cite{andrew}

The extended two-orbital Hubbard Hamiltonian 
is exactly investigated using the Lanczos algorithm\cite{dagottoRMP,lanczos} 
on a tilted $\sqrt{8}\times\sqrt{8}$ cluster, as done
in previous studies.~\cite{dagottoRMP,Daghofer:2008,moreo} In spite of the small
size of the cluster, this still requires 
substantial computational resources. More specifically, 
even exploiting the Hamiltonian symmetries the calculation of 
the undoped-limit ground state of the eight-sites cluster 
 still requires a basis with $\sim 2$-20\;M states 
(slightly more demanding than a 16-site cluster one-band Hubbard model), 
depending on the subspace explored. Runs applying the Lanczos technique had to be
performed for all the allowed momenta ${\bf k}$ of the cluster, and for all the 
quantum numbers under rotations and reflections 
(i.e. all the irreducible representations $A_{1g}$, $A_{2g}$, 
$B_{1g}$, $B_{2g}$, and $E_g$ 
of the $D_{4h}$  symmetry group~\cite{moreo}), 
and also for all the $z$-axis total spin projections. 
In addition, the computation of 
binding energies for the case of hole doping requires calculations for 
a number of electrons $N$ equal to 14, 15, and 16, varying 
$U$, $J_{\rm H}$, and $J_{\rm NN}$ 
using a fine grid. For these reasons, the overall effort amounted to $\sim 8,000$ 
diagonalizations of the cluster, supplemented also by calculations of dynamical 
properties, using a Penguin 128GB Altus 3600 computer.

\section{Results} \label{results}

It is important 
to remind the readers that the most commonly used 
models for the pnictides are
usually defined in the so-called unfolded 
Brillouin zone.\cite{raghu,moreo,graser,plee,our3} Due to the symmetry of the Fe-As 
planes,\cite{raghu,plee} it is possible to describe the pnictides using Fe-only
effective models where the As atoms merely provide a bridge for 
the electronic hopping between the irons. 
Under this approximation only one Fe atom is left 
per unit cell to describe these materials.\cite{foot}  
As a result of these considerations, the 
number of orbitals to be considered is reduced by half, which is a 
computational advantage, and the size of the Brillouin zone (BZ) is doubled.
For this reason, the momentum in the unfolded zone is dubbed ``pseudocrystal''
momentum.\cite{plee} 
In order to relate the model results to experiments addressing the BZ
corresponding to two Fe atoms, it 
is necessary to ``fold'' the extended BZ in such a way that the pseudocrystal 
momentum ${\bf k}=(\pi,\pi)$ is folded onto momentum $(0,0)$.
The 
physical difference between states with ${\bf k}=(0,0)$ and $(\pi,\pi)$ is that
the first indicates a bonding and the second an antibonding combination of
the $d$-orbitals in the two Fe atoms in the two-atoms unit cell.
In the presentation of our results below, ${\bf k}$ will stand for pseudocrystal
momentum.

\subsection{Phase diagram}

The relative symmetry between the undoped ($N=16$) ground state (GS) and 
the $N=14$ GS has been studied with the Lanczos technique varying $U/|t_1|$ and $J_{\rm H}/U$. 
The undoped GS was found to have momentum ${\bf k}=(0,0)$ and it
transforms according to the $A_{1g}$ representation of the $D_{4h}$ 
group, for all the values of $J_{\rm H}$ and $U$ studied here, 
in agreement with
previous results.~\cite{moreo} However, a surprising result 
found in the present study of the hole-doped extended 
two-orbital model is the presence of many competing 
low-energy states not only with different symmetries as in the electron doped
case,\cite{andrew} but also with different pseudocrystal momenta ${\bf k}$. In other
words, low
lying states with both ${\bf k}=(0,0)$ and $(\pi,\pi)$ were found in our Lanczos
investigation. 
This is compatible with previous mean-field approximation results that also reported 
low-energy spin-singlet pair states with momentum $(\pi,\pi)$.\cite{Gao10}

\begin{figure}[thbp]
\begin{center}
\includegraphics[width=8.5cm,clip,angle=0]{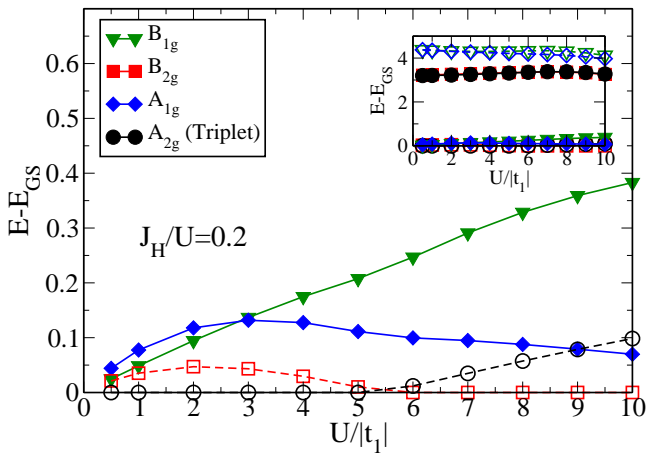}
\vskip -0.3cm
\caption{(Color online) Difference between the energy of the lowest excited state
with the symmetry and momentum indicated and the ground state.  Full (open) symbols denote 
${\bf k}=(0,0)$ (${\bf k}=(\pi,\pi)$). 
The results are obtained using the Lanczos algorithm for the two-orbital 
model in an eight-site cluster with 14 electrons (two holes doping), varying
the Hubbard repulsion $U$, and at a fixed $J_{\rm H}/U=0.20$. 
The inset shows a larger energy range
in which the lowest lying state with each symmetry is displayed. The results shown
in this figure are without the extra $J_{\rm NN}$ and $J_{\rm NNN}$ terms.}
\label{GS}
\end{center}
\end{figure}

The competition among low-lying states with different symmetries and with
different values
of ${\bf k}$ is presented in Fig.~\ref{GS} for the case $J_{\rm H}/U=0.20$, without
the extra ``$J$'' terms. 
Numerically, it was found that the ground state for 14 electrons
has crystal momentum $(\pi,\pi)$. For small values of $U$ this state is a 
triplet with $A_{2g}$ symmetry (open circles in the figure). 
With increasing $U$, a transition (via a level crossing) occurs  at
$U\sim 6|t_1|$ to a spin-singlet ground state with $B_{2g}$ symmetry 
(open squares in the figure). However, 
it can be observed that there are states with ${\bf k}=(0,0)$ that
have very similar energies. For example,
for this pseudocrystal momentum, and  in the weak coupling regime, 
a spin-singlet state with $B_{1g}$
symmetry (represented with filled triangles in the figure) 
is the closest in energy to the ground state, 
while for $U\ge 3|t_1|$ a spin-singlet state with $A_{1g}$ symmetry prevails
(represented with filled diamonds in the figure).

Similar results were found for all the values of $U$ and $J_{\rm H}$ studied, i.e., 
the $N=14$ ground state has total momentum 
${\bf k}=(\pi,\pi)$ but there are ${\bf k}=(0,0)$ states close in energy
with a different symmetry. For this reason, the phase diagrams obtained by varying 
$J_{\rm H}/U$ and $U/|t_1|$ 
for {\it both} values of the pseudocrystal momentum will be presented.
 
\begin{figure}[thbp]
\begin{center}
\includegraphics[width=8.5cm,clip,angle=0]{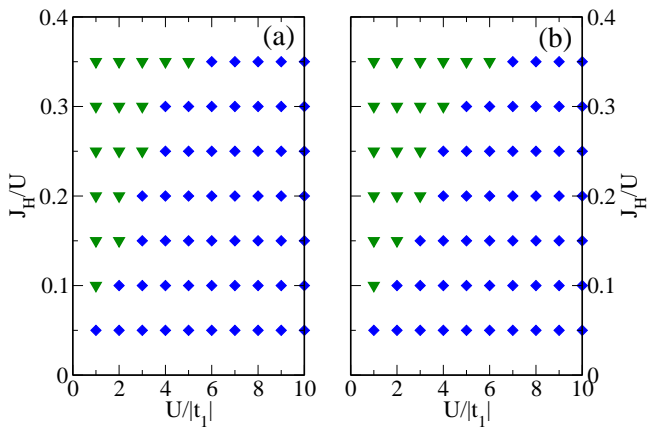}
\vskip -0.3cm
\caption{(Color online)  Relative symmetry between the $N$=16 (undoped) and 
$N$=14 (with ${\bf k}=(0,0)$) ground states, varying $U$ and $J_{\rm H}/U$.
Triangles denote $B_{1g}$-symmetric singlets, and diamonds $A_{1g}$-symmetric singlets.
(a) Results for couplings $J_{\rm NN}=J_{\rm NNN}=0$.  
(b) Results for the lowest value of ($J_{\rm NN},J_{\rm NNN}$) 
where binding appears with a fixed ratio $J_{\rm NN}/J_{\rm NNN}=0.93$.\cite{andrew}}
\label{Fig1}
\end{center}
\end{figure}

The relative symmetry between the ground state with two electrons less 
than half filling with total pseudocrystal momentum ${\bf k}=(0,0)$ and the 
undoped ground state is shown in Fig.~\ref{Fig1}(a), varying $J_{\rm H}/U$ 
and $U/|t_1|$. A region with symmetry $B_{1g}$, indicated by the  
triangles, is found for small $U/|t_1|$ (roughly $U/|t_1|\leq ~3$) and
moderate to large values of  $J_{\rm H}/U$. For larger values of $U/|t_1|$, 
the symmetry changes to $A_{1g}$.  
A similar transition from $B_{1g}$ to $A_{1g}$ (extended $s$-wave) has
been found using the RPA technique for an electron-doped five-orbital
model at $J_{\rm H}=0$.\cite{graser}
The binding energy $E_{\rm B}$, defined as
$E_{\rm B}=E(14)+E(16)-2E(15)$, where $E(N)$ is the GS
energy for $N$ electrons, was also calculated. It was found that without 
the addition of Heisenberg terms there are no regions with binding.

\begin{figure}[thbp]
\begin{center}
\includegraphics[width=8.5cm,clip,angle=0]{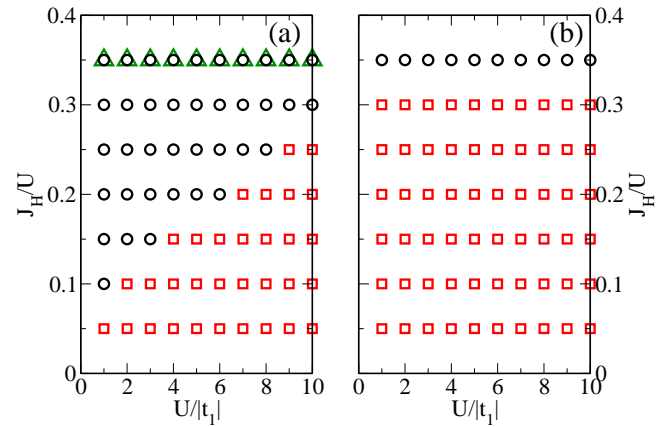}
\vskip -0.3cm
\caption{(Color online)  Relative symmetry between the $N$=16 (undoped) and 
$N$=14 (with ${\bf k}=(\pi,\pi)$) ground states 
 varying $U$ and $J_{\rm H}/U$.
Circles denote spin-triplet states and squares 
$B_{2g}$-symmetric singlets.
(a) Results for couplings $J_{\rm NN}=J_{\rm NNN}=0$. 
Open triangles indicate binding. 
(b) Results for the lowest value of ($J_{\rm NN},J_{\rm NNN}$) 
where binding appears with a fixed ratio $J_{\rm NN}/J_{\rm NNN}=0.93$.\cite{andrew}}
\label{Fig2}
\end{center}
\end{figure}

For the other case of a pseudocrystal momentum ${\bf k}=(\pi,\pi)$, the analogous numerical results are shown 
in Fig.~\ref{Fig2}(a). It was found that an $A_{2g}$ 
spin-triplet ground state, indicated by circles in the figure, 
dominates for large values of
$J_{\rm H}/U$ and small $U/|t_1|$.
For the electron-doped model, an $A_{2g}$ spin-triplet with momentum $(0,0)$
was similarly observed at large $J_{\rm H}$ and small $U$.\cite{Daghofer:2008,moreo}  
For smaller 
$J_{\rm H}/U$ and larger $U/|t_1|$, a spin-singlet ground state with 
$B_{2g}$ symmetry is the ground state.  For this pseudocrystal momentum, the binding 
energy was calculated as well: binding was obtained for $J_{\rm H}/U=0.35$ where a
spin-triplet ground state 
with symmetry $A_{2g}$ prevails (see open triangles in the
figure).

\subsection{Binding stabilization.}\label{binding} 

To stabilize hole binding in the two-orbital model, we will 
proceed as in the previous investigation of the electron-doped case\cite{andrew} 
by adding extra Heisenberg terms, namely
a NN coupling $J_{\rm NN}$ and a NNN coupling $J_{\rm NNN}$ as
discussed in Sec.~\ref{sec:model}. As in the electron
doped case, and as already explained, 
$J_{\rm NN}$ will be varied while $J_{\rm NN}/J_{\rm NNN}$
will be kept fixed at the value 0.93.\cite{andrew} 

The results for pseudocrystal  momentum ${\bf k}=(0,0)$ are presented in
Fig.~\ref{Fig1}(b), showing the symmetry of the hole-doped ground state 
for the lowest value of $J_{\rm NN}$ where binding of holes is achieved. 
The phase diagram remains largely unchanged by the addition of the Heisenberg 
terms except for the $B_{1g}$ region that has expanded slightly
towards larger values of $U$. 
On the other hand, for
states with momentum ${\bf k}=(\pi,\pi)$ the spin-triplet region
virtually disappears (Fig.~\ref{Fig2}(b)), except for those triplet
states that already had $E_B<0$ at $J_{\rm NN}=0$, 
leaving behind a much larger
$B_{2g}$ region in parameter space.

\begin{figure}[thbp]
\begin{center}
\vskip -0.65cm
\includegraphics[width=8.9cm,clip,angle=0]{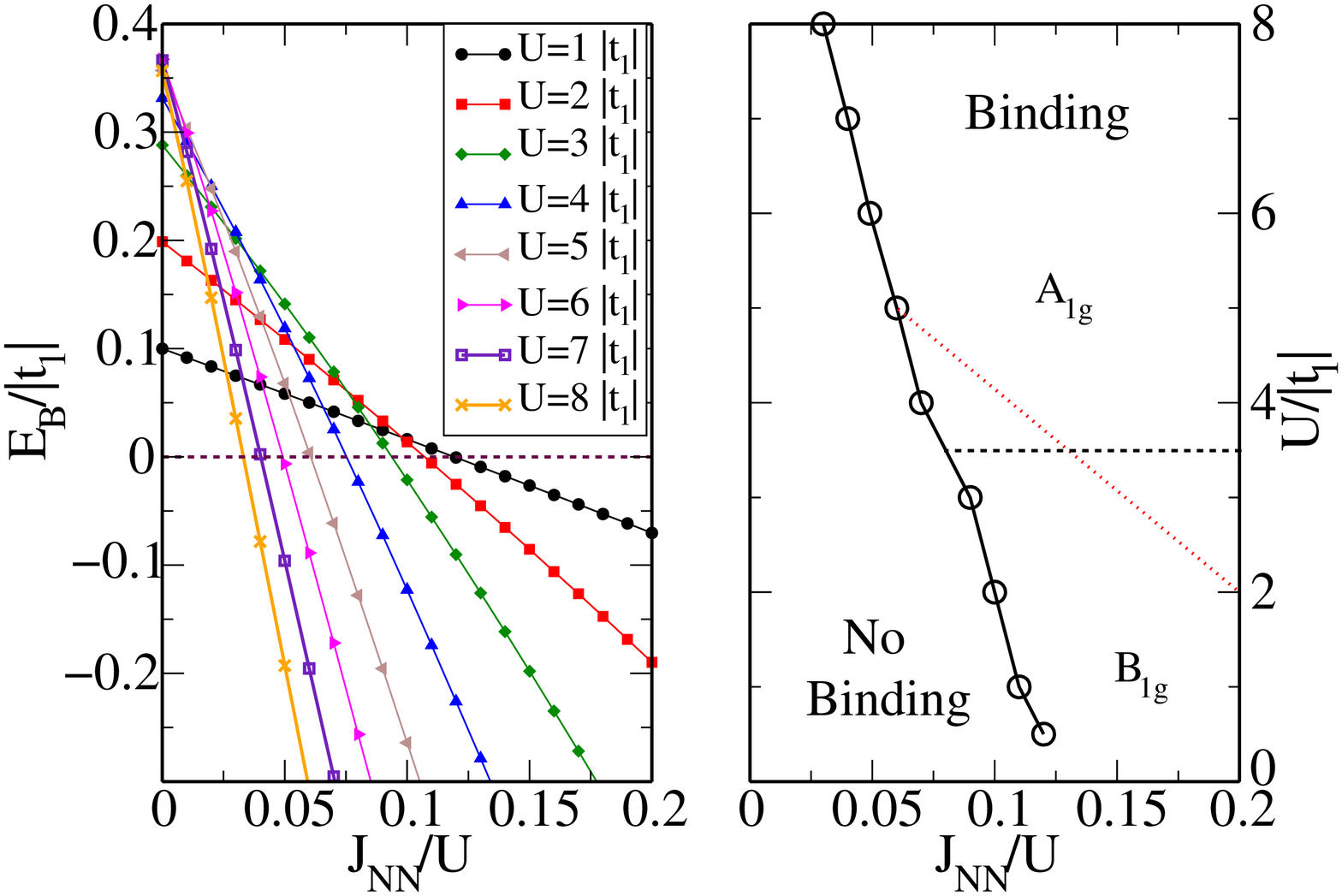}
\vskip -0.3cm
\caption{(Color online) Results for states with total momentum ${\bf k}=(0,0)$.
(a) $E_{\rm B}/|t_1|$ {\it vs.} 
$J_{\rm NN}/U$ for different 
values of $U/|t_1|$, at $J_{\rm H}/U=0.2$ and $J_{\rm NN}/J_{\rm NNN}=0.93$. 
(b) Phase diagram showing ``Binding'' and ``No Binding'' 
regions and the symmetry of the two-hole 
bound state varying
$U/|t_1|$ and $J_{\rm NN}/U$, at a fixed $J_{\rm H}/U=0.2$.  
The shaded area
indicates the so-called ``physical region'' obtained 
from standard mean-field calculations that were compared
with neutrons, transport, and photoemission experimental results.\cite{qinlong}
The doted line is for Fig.~\ref{fig:overlaps} (a).}
\label{Fig3}
\end{center}
\end{figure}

\begin{figure}[thbp]
\begin{center}
\vskip -0.55cm
\includegraphics[width=8.9cm,clip,angle=0]{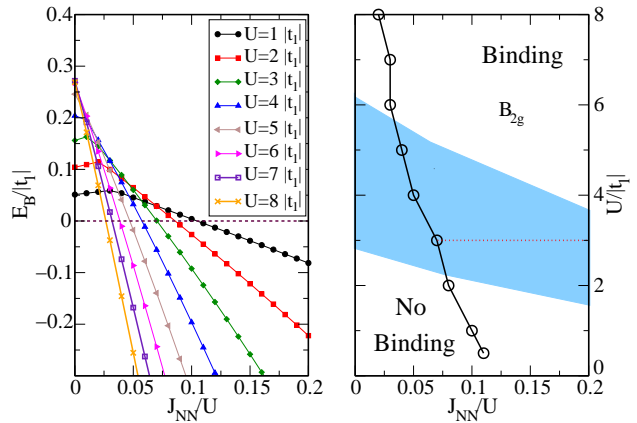}
\vskip -0.3cm
\caption{(Color online) Results for states with total momentum ${\bf k}=(\pi,\pi)$. 
(a) $E_{\rm B}/|t_1|$ {\it vs.} 
$J_{\rm NN}/U$ for different 
values of $U/|t_1|$, at $J_{\rm H}/U=0.2$ and $J_{\rm NN}/J_{\rm NNN}=0.93$. 
(b) Phase diagram showing ``Binding'' and ``No Binding'' 
regions and the symmetry of the two-hole 
bound state varying
$U/|t_1|$ and $J_{\rm NN}/U$, at a fixed $J_{\rm H}/U=0.2$. The shaded region 
indicates the ``physical region'' according to standard mean-field 
calculations.\cite{qinlong} 
The doted line is for Fig.~\ref{fig:overlaps}(b).}
\label{Fig4}
\end{center}
\end{figure}

In Fig.~\ref{Fig3}(a), the binding energy $E_B$ {\it vs}. $J_{\rm NN}/U$ 
for states with momentum ${\bf k}=(0,0)$ is shown
for several values of $U$ and at a fixed (realistic) $J_{\rm H}/U=0.2$. 
Increasing $J_{\rm NN}$ eventually induces binding for 
all $U$'s. The value of $J_{\rm NN}/U$ where binding 
occurs decreases as $U$ increases. Figure~\ref{Fig4}(a) shows the same
information but for the states with  momentum 
${\bf k}=(\pi,\pi)$, where a similar qualitative behavior is observed.

\vskip 0.3cm

A study of the binding energy 
$E_{\rm B}$ and the relative symmetry between the $N$=16 and 
14 GS's  allows us to construct phase diagrams in 
the $(U,J_{\rm NN}/U)$ plane.
In Fig.~\ref{Fig3}(b), typical
results for the case $J_{\rm H}/U=0.2$ are shown~\cite{foot1} for the states
with total momentum ${\bf k}=(0,0)$. The bound state has $A_{1g}$ symmetry
in most of the binding region, but a state with $B_{1g}$ symmetry 
prevails at smaller $U$ values ($\sim 3|t_1|$). In Fig.~\ref{Fig4}(b)
the same information is displayed but for states with total momentum
${\bf k}=(\pi,\pi)$.  In this case, the entire binding region, except 
for $J_{\rm H}/U>0.3$,  has
$B_{2g}$ symmetry.  All of the above symmetries appear inside the proper 
magnetic/metallic region of the undoped limit 
(indicated with shading in the figures) that were obtained in previous
mean-field calculations~\cite{rong,qinlong} extended to incorporate $J_{\rm
  NN}$.\cite{andrew} 

\vskip0.3cm

\subsection{Magnetism}\label{sec:sk}

Since the two-orbital Hubbard model for the pnictides is not particle-hole symmetric, 
it is interesting to study how the nature of the doping, namely electrons vs. holes,
affects the intensity of the magnetic order. In the actual materials, 
experimental results have shown that the in-plane resistivity of electron
and hole-doped FeAs-based pnictides displays a larger anisotropy in the 
electron-doped case.\cite{ying}  Thus, it has been conjectured that the $xz$/$yz$ magnetism
is stronger in the electron-doped case, while in the hole-doped case it is
weaker with a growing contribution of 
the $xy$ orbital, disregarded in the two-orbital 
model, that forms the hole pocket around $M$.\cite{ying} A similar conclusion 
was reached  via the FLEX approximation for the case of electron and hole 
doping of a five-orbital Hubbard model.\cite{ikeda}  

\vskip1.5cm
\begin{figure}[thbp]
\begin{center}
\includegraphics[width=8.5cm,clip,angle=0]{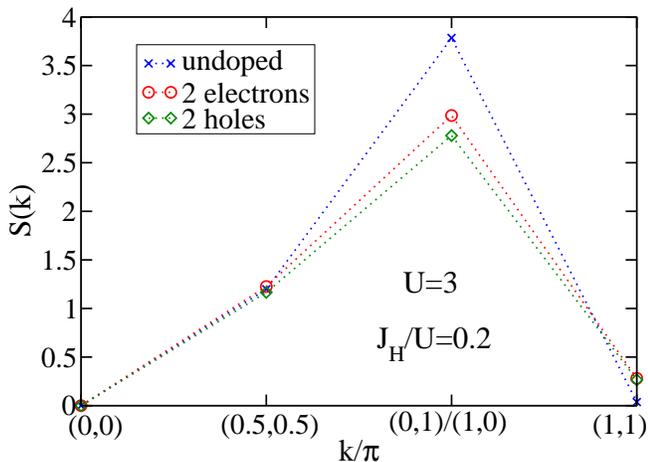}
\vskip -0.3cm
\caption{(Color online) Numerically calculated 
magnetic structure factor $S({\bf k})$, as a function
of the momentum, using an eight sites cluster. 
Results for the undoped $N$=16, electron-doped $N$=18, and hole-doped $N$=14
cases are indicated, for couplings $U/|t_1|=3$, $J_{\rm H}/U=0.2$, and $J_{\rm NN}/U=0.2$.}
\label{Figsk}
\end{center}
\end{figure}

The results for the two-orbital model studied here are shown in Fig.~\ref{Figsk} where the 
magnetic structure factor $S({\bf k})$ is shown in the undoped (crosses), 
electron doped (circles), and hole-doped (diamonds) regimes, at fixed couplings
$U=3$, $J_{\rm H}/U=0.2$, and
$J_{\rm NN}/U=0.2$, namely in the mean-field calculated ``physical region'' 
indicated in Fig.~\ref{Fig4}. While doping reduces the strength of the peak
at ${\bf k}=(\pi,0)$, it is interesting to notice that the intensity is slightly more 
reduced in the hole-doped case. These results lend qualitative support to the 
notion that the magnetism in the $xz$ and $yz$ orbitals is stronger in the
electron-doped case, and it becomes reduced when holes are introduced.

\subsection{Overlap Integrals}\label{sec:overlaps}

In this subsection, the functional forms of the hole pairing operators that produce the
hole bound states will be  analyzed.
With this goal, the overlap defined by
\begin{equation}\label{overlap}
\langle\Psi_{(N=14)}({\bf k'})|\Delta_{{\bf k'-k},i}|\Psi_{(N=16)}({\bf k}=(0,0))\rangle
\end{equation}
\noindent was calculated  using the Lanczos algorithm 
along the paths indicated by the 
dotted lines in the phase 
diagrams shown in panels (b) of Figs.~\ref{Fig3} and ~\ref{Fig4}.
Notice that for $|\Psi_{14}({\bf k'})\rangle$ the pseudocrystal momentum ${\bf k'}$
will take the values $(0,0)$ and $(\pi,\pi)$ and, thus, a pairing operator with
the appropriate ${\bf k'-k}$ has to be used to ensure a non-zero overlap.  
The ground state $|\Psi(N)\rangle$ in the subspace of $N$ 
electrons was used, and the operator in Eq.~(3) was defined as
\begin{equation}\label{delta} 
\Delta_{{\bf k},i}=\sum_{\alpha \beta} f({\bf k})(\sigma_i)_{\alpha \beta}
d_{{\bf k},\alpha,\uparrow}d_{{\bf k},\beta,\downarrow},
\end{equation}
\noindent where $d_{{\bf k},\alpha,\sigma}$ destroys
an electron with spin $z$-axis projection $\sigma$, 
at orbital $\alpha=x,y$, and with momentum ${\bf k}$.
The structure factor $f({\bf k})$ arises from the spatial 
location of the fermions forming the pair,~\cite{moreo}
and $\sigma_i$ are the 
Pauli matrices ($i=1,2,3$) or the $2\times 2$ identity matrix $\sigma_0$ 
($i=0$). Note that $\sigma_1$ and $\sigma_2$ imply an inter-orbital pairing. 
Overlaps for
all the symmetries in Ref.~\onlinecite{moreo}, 
and with NN and NNN locations for the electronic pairs, 
were numerically evaluated.

\begin{figure}[t!]
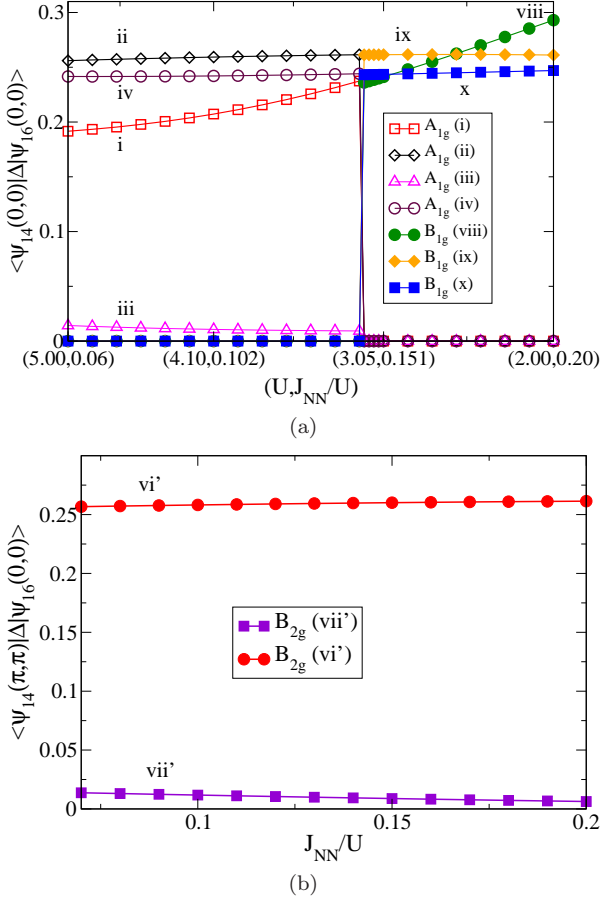

\begin{center}
\subfigure[]{\includegraphics[width=7.9cm,clip,angle=0]{./K00_overlap.eps}}
\subfigure[]{\includegraphics[width=7.9cm,clip,angle=0]{./KPIPI_overlap.eps}}
\vskip -0.3cm
\caption{(Color online) Overlap 
$\langle\Psi(N=14)|\Delta_{{\bf k},i}|\Psi(N=16)\rangle$ {\it vs.} 
$J_{\rm NN}/U$ for 
the indicated pairing 
operators, at $U=3$~$|t_1|$ and $J_{\rm H}/U=0.2$, for (a) states with 
total momentum ${\bf
    k}=(0,0)$ along the dotted path in Fig.~\ref{Fig3}(b), and  (b) states with total momentum ${\bf
    k}=(\pi,\pi)$ along the dotted path in Fig.~\ref{Fig4}(b).}
\label{fig:overlaps}
\end{center}
\end{figure}

In Fig.~\ref{fig:overlaps}(a), the overlaps for pairing operators with 
pseudocrystal momentum ${\bf k}=(0,0)$ are presented for values of $U$ and
$J_{\rm H}$ along the dotted path in Fig.~\ref{Fig3}(b).
In the $A_{1g}$ region in
Fig.~\ref{fig:overlaps}(a), we found that the same four pairing
operators that have a finite overlap in the electron doped case
\cite{andrew} also have one here. However, the relative strength of
the overlaps differ. For consistency, we will use 
the same labeling for the operators as in Ref.~\onlinecite{andrew}.  
The $A_{1g}$ operator with the largest overlap is the operator (ii), 
i.e. the $s_{\pm}$ operator characterized by $f({\bf k})\sigma_i =
(\cos k_x\cos k_y)\sigma_0$, as in the electron doped case;\cite{andrew} 
it is indicated by
hollow diamonds in Fig.~\ref{fig:overlaps}(a). However, in the 
hole-doped system the overlap 
for the pairing operator (iv)
characterized by $(\cos k_x-\cos k_y)\sigma_3$ (hollow 
circles) follows in strength; this operator had the weakest overlap in
the electron doped case.\cite{andrew} The pairing operator (i) with   
$(\cos k_x+\cos k_y)\sigma_0$ (hollow squares) has an 
overlap almost as strong as in the electron-doped case. Finally, the overlap
corresponding to the 
operator (iii)$(\sin k_x\sin k_y)\sigma_1$ (hollow triangles) is even 
more suppressed upon hole doping than upon electron doping.

In the region where the pairs have $B_{1g}$ symmetry 
there are three pairing operators with
large overlaps : (viii) $(\cos k_x+\cos k_y)\sigma_3$ (solid
circles); (ix) $(\cos k_x\cos k_y)\sigma_3$ (solid
diamonds); and (x) $(\cos k_x-\cos k_y)\sigma_0$ (solid
squares).  At small values of  $J_{\rm NN}/U$, (ix) has the largest overlap
amplitude followed by (x) and (viii).  However, as
$J_{\rm NN}/U$ increases (viii) overtakes (ix).

For the case of pairing operators with pseudocrystal momentum ${\bf k}=(\pi,\pi)$, there
is one contribution that clearly dominates, see Fig.~\ref{fig:overlaps}(b): (vi') $(\cos
k_x\cos k_y)\sigma_1$ which corresponds to a NNN pair with 
$B_{2g}$ symmetry. The prime in the label is used 
to remind the reader that the operator has a different
pseudocrystal momentum from the  $B_{2g}$ state with the same label
discussed in the electron-doped case.\cite{andrew} The only other nonzero pairing
overlap occurs for (vii') $(\sin k_x\sin k_y)\sigma_0$ and has a much
smaller amplitude than (vi'). Interestingly, the nearest-neighbor
$B_{2g}$ operator (v) characterized 
by $(\cos k_x+\cos k_y)\sigma_1$ that had the strongest overlap in the electron 
doped case\cite{andrew} has zero overlap in the case studied in this manuscript.
All the gaps for the pairing operators with B$_{2g}$ symmetry have nodes along the $x$ and $y$ axes.

\begin{figure}[thbp]
\begin{center}
\includegraphics[width=8.5cm,clip,angle=0]{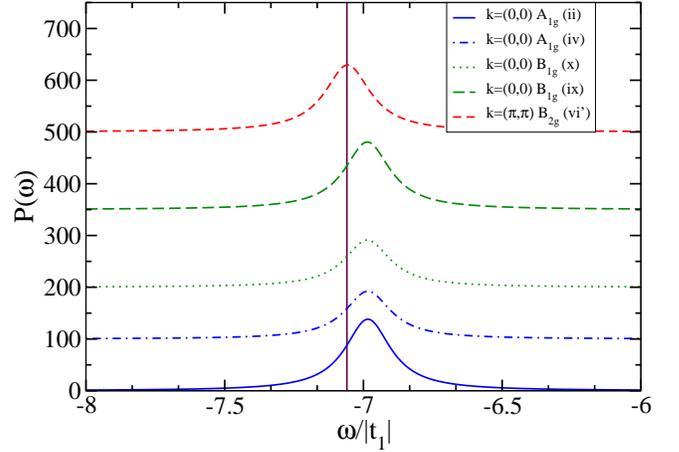}
\caption{(Color online) Dynamic pairing 
susceptibility for the pairing operators with total momentum ${\bf
    k}=(\pi,\pi)$ (operators with $B_{2g}$ symmetry) and with  total momentum ${\bf k}=(0,0)$ (operators with $B_{1g}$ and $A_{1g}$  symmetry)  (see text), at
$U=3.0$~$|t_1|$, $J_{\rm H}/U=0.2$, and $J_{\rm NN}/U=0.10$. 
The vertical line indicates $E_{GS}(14)-E_{GS}(16)$.}
\label{dyncombined}
\end{center}
\end{figure}

\subsection{ Dynamical Pair Susceptibilities.}\label{dynamical}

 To complete our analysis, 
the dynamical pair susceptibilities defined as
\begin{equation}\label{eg:Pomega}
P(\omega)=\int_{-\infty}^{\infty}dt e^{i\omega t}\langle\Delta^{\dagger}_{{\bf k},i}(t)
\Delta_{{\bf k},i}(0)\rangle,
\end{equation}
 were also studied in the state with $N=16$ 
for the pairing operators $\Delta_{{\bf k},i}$ 
introduced in Sec.~\ref{sec:overlaps}. Notice that the calculated spectral decomposition 
involves excited states with $N=14$.
The procedure described in Ref.~\onlinecite{jose} 
in the context of the cuprates will be followed. As discussed above,
for $N=14$ there are several low-energy states near the ground state 
that have different symmetries. 
The dynamical pair susceptibilities show that most of these low-lying states
have a large overlap with $\Delta_{{\bf k},i}|\Psi_{\rm N=16}(0)\rangle$ for  
$\Delta_{{\bf k},i}$ with the appropriate symmetry. 
In Fig.~\ref{dyncombined}, results for $U=3|t_1|$, $J_{\rm H}/U=0.2$, 
and $J_{\rm NN}/U=0.10$ are presented. Large overlaps with low-lying $N=14$ states
are observed for operators (ii) and (iv) with $A_{1g}$ symmetry and (viii) and (x) with $B_{1g}$ symmetry, 
as well as for operator (vi') with $B_{2g}$
symmetry and pseudocrystal momentum ${\bf k}=(\pi,\pi)$. 

It is interesting to compare the results obtained for the dynamical
pair susceptibility upon hole doping with those
obtained for electron doping.\cite{andrew} In both cases, large
susceptibilities for the low-lying states with $A_{1g}$, $B_{1g}$, and
$B_{2g}$ symmetries are found. This is remarkably different from the
case of models for the cuprates where an analogous low-lying overlap analysis
showed that $d_{x^2-y^2}$-wave symmetry clearly dominates over all others.\cite{jose} 

Returning to pnictides, a similarity
between the electron and hole-doped cases is that the $B_{1g}$
pairing operator that has the highest 
susceptibility, state (viii), is different from the $B_{1g}$ pairing state
for the cuprates.
It corresponds to Cooper pairs mainly located on NNN sites, as opposed to the
dominant NN contribution in the cuprates, and in the orbital basis used
here the $B_{1g}$ symmetry is realized by the orbital degree of freedom.
In addition, the susceptibilities indicate that while NN pairs are favored in
the electron doped case, NNN have larger susceptibilities upon hole
doping for $A_{1g}$ and $B_{2g}$ symmetries.

\section{ Conclusions}\label{conclusions} 

The properties of a recently introduced two-orbital extended 
Hubbard model for the pnictides have been studied
upon hole doping with the help of the Lanczos method. The results were 
contrasted with the previously studied electron-doped case.\cite{moreo,andrew}
Due to the lack of particle-hole symmetry in the Hamiltonian, the results, as
expected, are quantitatively different in both cases. However, an additional surprising
characteristic of the hole-doped ground state is that it has pseudocrystal momentum 
$(\pi,\pi)$. In the reduced Brillouin zone representation corresponding to the
physical two Fe-atoms per unit cell description of the pnictides, having a 
nonzero pseudocrystal momentum means that
the ground state is characterized by antibonding rather than bonding 
combinations of the orbitals of the two Fe atoms in the unit cell. In terms of 
the pairing operators that are favored, 
it means that the pairs would arise from
hole carriers located at the hole pockets at $\Gamma$ and at $M$ 
in the unfolded Brillouin 
zone. Interestingly, the five-orbital model 
for the pnictides\cite{kemper} shows that 
upon hole doping a hole pocket, absent in the electron-doped case, develops
around $M$ and the role of this pocket plays an important role in the 
properties of the hole-doped materials.\cite{kemper,ikeda} Our results may indicate 
that a simple toy model, such as the two-orbital model, 
could be used to study the role 
that a hole-pocket at $M$ plays when multiorbital Hubbard models are 
hole-doped.

In spite of this difference in the pseudocrystal momentum quantum number, 
there are several commonalities between the hole- and electron-doped
two-orbital Hubbard models. The most important feature is that there are 
several low-lying states with different symmetries 
close to the undoped ground state. For 
this reason, the symmetry of the doped states is strongly dependent on the
actual values of the interaction parameters. Spin-singlet 
states that transform according
to the irreducible representations $A_{1g}$, $B_{1g}$, and $B_{2g}$ were
obtained both for hole and for electron doping. The richness of the phase 
diagrams unveiled here, and in the cited previous investigations, 
suggests that the symmetry of the pairing state in the pnictides is 
likely to depend on the material as well as on the type of doped carriers 
(electrons or holes) and on the density of dopants.

\section{Acknowledgments.} 

This work was supported by 
the U.S. Department of Energy, Office of Basic Energy Sciences,
Materials Sciences and Engineering Division, and also by the National Science 
Foundation under grant DMR-11-04386 (A.N., W.G., A.M., 
E.D.), CONICET, Argentina (J.R.), and by the DFG under the Emmy-Noether 
program (M.D.).


\begin{thebibliography}{10}
\bibitem{foot1} At $J_{\rm H}/U=0.3$ $(0.1)$ the phase diagram 
Fig.~\ref{Fig2}(b) was found to be
qualitatively similar to the  case of $J_{\rm H}/U=0.2$ except that the 
``No Binding'' region shrinks (expands) in size.
\end{thebibliography}
\end{document}